%% file: IEEEPaper.tex
\documentclass[conference]{IEEEtran}
\IEEEoverridecommandlockouts
\usepackage{cite}
\usepackage{amsmath,amssymb,amsfonts}
\usepackage{graphicx}
\usepackage{textcomp}
\usepackage{xcolor}

\usepackage[inline]{enumitem}

\usepackage{color}

\definecolor{highlight}{RGB}{255,127,127}

\usepackage{balance}

\usepackage[noend]{algpseudocode}
\usepackage[boxed]{algorithm}
\usepackage[mode=buildnew]{standalone}

\usepackage{graphicx}
\usepackage{caption}
\usepackage{subcaption}
\usepackage{epstopdf}
\usepackage[mode=buildnew]{standalone}

\usepackage{mathtools}
\usepackage{amssymb}

\usepackage{pgfplots}
\usepackage{xcolor}
\usepgfplotslibrary{colorbrewer}
\pgfplotsset{cycle list/Dark2}
\usetikzlibrary{matrix}
\pgfplotsset{compat=newest}
\usepgfplotslibrary{groupplots}

\usepackage{multirow}
\renewcommand{\arraystretch}{1.4}

\def\BibTeX{{\rm B\kern-.05em{\sc i\kern-.025em b}\kern-.08em
    T\kern-.1667em\lower.7ex\hbox{E}\kern-.125emX}}

\begin{document}
	\title{Bounded Dijkstra (BD): Search Space Reduction \\ for Expediting Shortest Path Subroutines}
%-- better SP/SPT impact?
%- eval of BD with A*?
%- explain better BF (depending on the BF extension used)?
%- topology dependence

	\author{\IEEEauthorblockN{Amaury Van Bemten, Jochen W. Guck, Carmen Mas Machuca and Wolfgang Kellerer}
		\IEEEauthorblockA{Lehrstuhl f\"ur Kommunikationsnetze\\Technical University of Munich \\
			Email: \{amaury.van-bemten, guck, cmas, wolfgang.kellerer\}@tum.de}}
	
	\maketitle
	
	\begin{abstract}
		The \emph{shortest path} (SP) and \emph{shortest paths tree} (SPT) problems arise both as direct applications and as subroutines of overlay algorithms solving more complex problems such as the \emph{constrained shortest path} (CSP) or the \emph{constrained minimum Steiner tree} (CMST) problems.
		Often, such algorithms do not use the result of an SP subroutine if its total cost is greater than a given bound.
		For example, for delay-constrained problems, paths resulting from a least-delay SP run and whose delay is greater than the delay constraint of the original problem are not used by the overlay algorithm to construct its solution.
		As a result of the existence of these bounds, and because the Dijkstra SP algorithm discovers paths in increasing order of cost, we can terminate the SP search earlier, i.e., once it is known that paths with a greater total cost will not be considered by the overlay algorithm.
		This early termination allows to reduce the runtime of the SP subroutine, thereby reducing the runtime of the overlay algorithm without impacting its final result.
		We refer to this adaptation of Dijkstra for centralized implementations as \emph{bounded Dijkstra} (BD).
		On the example of CSP algorithms, we confirm the usefulness of BD by showing that it can reduce the runtime of some algorithms by 75\% on average.
	\end{abstract}
	

\IEEEpeerreviewmaketitle

\section{Introduction}

The \emph{shortest path} (SP) and \emph{shortest paths tree} (SPT) routing problems arise in a wide range of practical problems, both as direct applications~\cite{ahuja1993network} and as subroutine of other more complex problems such as the \emph{(multi-)constrained shortest path} (CSP and MCSP) and \emph{multi-constrained path} (MCP) routing problems~\cite{guck2017unicast}, which are encountered, e.g., when provisioning quality of service (QoS) in software-defined networks~\cite{guck2017detserv, van2018larac, zoppi2018achieving}.
When used as a subroutine, the result of one or several SP/SPT search(es) is used to determine a solution to the original problem.
For example, DCUR~\cite{salama1997distributed, reeves2000distributed} combines the least-delay and least-cost SPTs from all nodes to a given destination in order to compute a \emph{delay-constrained least-cost} (DCLC) path, i.e., a CSP, from a single source to a single destination.
As a result, optimizing SP/SPT procedures is an important issue, as it allows to reduce the runtime of the wide range of routing algorithms using them, those of which can have considerable runtime footprint~\cite{guck2017unicast, van2018routing}.

\begin{figure}
	\centering
	\includestandalone[width=\linewidth]{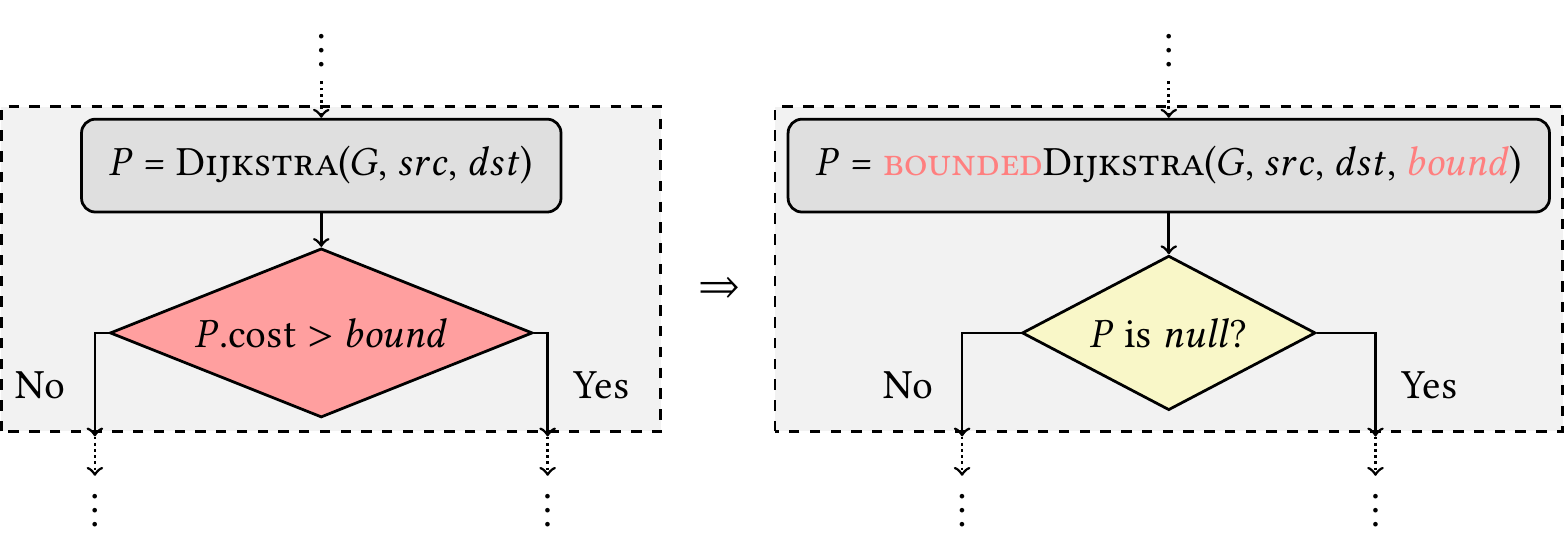}
	\vskip -0.1cm
	\caption{
		When using a shortest path subroutine, algorithms often do not use the paths returned by Dijkstra if they are more costly than a given \emph{bound}.
		The usage of bounded Dijkstra (BD) consists in incorporating this decision in the Dijkstra subroutine in order to avoid exploring these useless paths, thereby reducing the search space of Dijkstra. 
		The input and output of the overlay algorithm are left unchanged, only its runtime is affected.
	}\label{fig:jd-representation}
\end{figure}

For SP/SPT problems, the Dijkstra algorithm~\cite{dijkstra1959note} is commonly acknowledged as the fastest optimal algorithm for centralized implementations~\cite{cherkassky1996shortest}.
When using an SP/SPT algorithm as a subroutine, its result is often not used if its total cost is greater than a given bound.
For example, DCUR does not use the paths of its least-delay SPT search whose delay is greater than the delay constraint of the CSP problem.
Indeed, these paths cannot lead to a feasible solution of the original problem.
As a result, because Dijkstra discovers paths in increasing order of cost, the least-delay SPT search of DCUR can be stopped when paths with a delay higher than the delay constraint of the CSP problem are reached.
This early termination of Dijkstra allows to reduce the runtime of the SP/SPT search, thereby reducing the runtime of the overlay algorithm (e.g., DCUR) without impacting its result.
We refer to this simple adaptation of Dijkstra for centralized implementations as \emph{bounded Dijkstra} (BD). %\footnote{Because it only works when bounded together with an overlay algorithm.}. 
%This constitutes the main contribution of this paper.
BD can be used by any routing algorithm making use of one or several SP/SPT search(es) and able to provide a bound to these subroutines (Fig.~\ref{fig:jd-representation}). % above which results are unnecessary.

First, in Sec.~\ref{sec:related}, we present related work aiming at optimizing the runtime of SP/SPT searches.
Second, after presenting the simple functioning and the benefits of BD in Sec.~\ref{sec:jd}, we show how BD can be used, i.e., how a bound can be provided to the SP/SPT subroutines, in the particular case of centralized CSP algorithms (Sec.~\ref{sec:csp}).
We show that BD can be used by a wide range of algorithms.
Indeed, 20 out of 26 recently surveyed CSP algorithms~\cite{guck2017unicast} can make use of BD to improve their runtime.
For each of them, we detail how bounds can be provided to the SP/SPT subroutines.
Finally, in Sec.~\ref{sec:evaluation}, we evaluate the impact of BD on the performance of all these CSP algorithms.
We observe that BD can reduce the runtime of some CSP algorithms by 75\% on average.
For favorable cases, BD reduces the runtime of several algorithms by 96\% on average.
We further confirm that using BD does not change the final solution found by the algorithms.
These algorithms hence have only benefits in using BD: reduced runtime at no cost.
Due to the high number of algorithms, we only present the most interesting and insightful results and conclusions.
The entire set of raw results and graphs is publicly available at \textit{https://lora.lkn.ei.tum.de}~\cite{heaven-page}.

\section{Related Work}\label{sec:related}

The SP/SPT problem has been thoroughly investigated in the literature.
In this section, we classify the attempts at making SP/SPT routines faster in six categories for which we list representative examples and with respect to which we highlight our contribution.

\subsubsection{Heuristics}
Some approaches improve the runtime of SP/SPT searches by accepting to find sub-optimal solutions~\cite{fu2006heuristic}.
In contrast, for positive metrics, BD guarantees to find the optimal result.

\subsubsection{Hierarchical Routing}
A way of reducing the complexity of SP/SPT routing is to apply hierarchical routing, thereby running SP/SPT algorithms on smaller graphs~\cite{kleinrock1977hierarchical, jagadeesh2002heuristic}.
As it does not modify the subject graph, BD can be used as part of any such hierarchical routing scheme.

\subsubsection{Data Structure Optimizations}\label{sec:ds}
Several studies propose optimized data structures for the implementation of Dijkstra~\cite{fredman1987fibonacci, cherkassky1999buckets, thorup2003integer}.
BD is independent of the data structure used and can hence be used with any of these data structures.

\subsubsection{Improvements of Existing Algorithms}
Several proposals introduce extensions to the well-known Dijkstra~\cite{hart1968formal, goldberg2005computing} and Bellman-Ford~\cite{yen1970algorithm, bannister2012randomized} algorithms.
BD falls into this category but can be used in parallel with these improvements.

\subsubsection{Bi-Directional Searches}
Dijkstra explores the graph from the source towards the destination(s).
For SP problems, bi-directional searches, starting from both the source and the destination simultaneously, have been proposed~\cite{kwa1989bs, ikeda1994fast, rice2012bidirectional}, potentially pruning parts of the individual searches when possible~\cite{kwa1989bs}.
While bi-directional searches can only be used for SP problems, BD can be used for both SP and SPT problems.

\subsubsection{Preprocessing}
In the context of very large graphs such as road networks, a plethora of work~\cite{bast2016route} proposes to perform a preprocessing step computing intermediate information by, e.g., clustering nodes~\cite{hilger2009fast}, defining overlay graphs~\cite{holzer2009engineering}, defining important transit nodes~\cite{akiba2013fast}, or computing virtual links~\cite{abraham2012hierarchical}.
This preprocessed information is later used to solve routing requests faster.
The precomputation step being costly, these approaches are only suitable for solving a batch of requests on the same static topology.
Besides, these algorithms work well for large topologies but hardly outperform Dijkstra for general topologies~\cite{bast2016route}.
On the other hand, BD can provide significant benefit on general and dynamic topologies.
Some preprocessing algorithms involve SPT subroutines~\cite{akiba2013fast} and expedite these subroutines by pruning parts of the network because the corresponding information was already obtained from previous SPT runs.
In contrast, BD does not exploit previous searches in order to improve runtime but rather information provided by an overlay algorithm.
Further, these proposals simply prune parts of the network when a given condition is met, while BD completely terminates.

We note that bounded Dijkstra runs were already used in the literature~\cite{bose2008computing, lingas2009efficient, bourdelles2012routing, rice2013parameterized, varone2014many, varone2014insertion, simari2014fast}, but only as part of the design of new specific algorithms. 
Our contribution consists in the formalization and generalization of such an approach for any generic algorithm using SP/SPT subroutines, and in the quantification of its benefits for these algorithms.

\section{Bounded Dijkstra (BD)}\label{sec:jd}

In this section, we present the \emph{bounded Dijkstra} (BD) algorithm.
After presenting the context in which BD can be used (Sec.~\ref{sec:jd-context}), we describe the simple idea of the algorithm (Sec.~\ref{sec:jd-idea}) and detail the impact it can have on SP (Sec.~\ref{sec:sp}) and SPT (Sec.~\ref{sec:sp-tree}) searches.
Then, we show that the same idea can also be applied to the \emph{Bellman-Ford} (BF)~\cite{ford1956network, bellman1958routing} and  \emph{Chong}~\cite{chong1995finding} algorithms (Sec.~\ref{sec:bf} and \ref{sec:chong}), respectively another SP/SPT algorithm and a kSP/kSPT algorithm.

\subsection{Context: Centralized Bounded SP/SPT Subroutines}\label{sec:jd-context}

The \emph{shortest path} (SP) and \emph{shortest paths tree} (SPT) problems are core networking problems.
Besides in their direct applications, these problems are often encountered as subproblems of other more complex centralized problem settings.
For example, many \emph{(multi-)constrained shortest path} algorithms (CSP and MCSP) use results of SP/SPT searches to determine a solution to their problem~\cite{guck2017unicast}.
Similarly, \emph{multi-constrained path} (MCP) algorithms such as H\_MCP~\cite{korkmaz2001multi} or \emph{constrained minimum Steiner tree} (CMST) algorithms such as BSMA~\cite{zhu1995source} make use of an underlying SPT algorithm to construct a solution to their problem.
When an SP/SPT algorithm is used as such a subroutine of a centralized algorithm, it often happens that paths with a total cost greater than a given bound are not used.
For an SP search, this means that the result itself is not used.
For an SPT search, this means that the paths to some destinations (too costly) are not considered, while others are.
For example, for delay-constrained problems, paths resulting from a least-delay SP/SPT run which have a delay higher than the delay constraint are not considered.
Similarly, for \emph{delay-constrained least-cost} (DCLC), or CSP, problems, paths resulting from a least-cost SPT run which have a cost higher than the cost of the least-delay path will not be used, as these paths have a higher delay \emph{and} cost than the least-delay path.

\subsection{Idea: Early Termination for Search Space Reduction}\label{sec:jd-idea}

As a result of the existence of these bounds, and because the Dijkstra algorithm~\cite{dijkstra1959note} discovers paths in increasing order of cost, the SP/SPT searches can be stopped earlier, i.e., once it is known that paths with a greater total cost will not be considered by the overlay algorithm, thereby reducing the search space of Dijsktra and hence the runtime of the overlay algorithm.
We refer to such a modified version of Dijkstra as \emph{bounded Dijkstra} (BD).
The pseudo-code of BD is shown in Fig.~\ref{pseudo-code} and the required modification in the overlay algorithm is shown in Fig.~\ref{fig:jd-representation}.
For positive metrics, BD does not influence the result of the overlay algorithm, as the latter considers having no path and having a too costly path identically.
In order to use BD, an algorithm must of course be able to provide a bound value above which results are unnecessary.
We detail in Sec.~\ref{sec:csp}, on the example of CSP algorithms, how overlay algorithms can provide such bounds.

\subsection{Impact on an SP Search}\label{sec:sp}

\begin{figure}
	\footnotesize
	\begin{algorithmic}[1]
		\Function{\textcolor{highlight}{bounded}Dijkstra}{$G$, $src$, $dst$, \textcolor{highlight}{$bound$}}
		\State{Create empty priority queue $Q$}
		\For{\textbf{each} $node$ $\in$ $G$}
		\State{$node$.cost $\gets +\infty$}
		\EndFor
		
		\State{$src$.cost $\gets 0$}
		\State{$Q$.add($src$)}	
		\While{\textbf{not} $Q$.empty}
		\State{$node \gets Q$.popLeastCostNode()}\label{popping}
		\State{\textbf{if} $node$ \textbf{is} $dst$ \textbf{then} \Return \textsc{getPredecessors}($dst$)}\label{single-destination}
		\State{\textbf{if} $node$.visited \textbf{then} \textbf{continue}}
		\State{$node$.visited $\gets$ \textsc{true}}
		\For{\textbf{each} outgoing edge of $node$ \textbf{as} $edge$}
		\State{$newCost \gets node$.cost $+$ $edge.$cost}
		\textcolor{highlight}{\If{$newCost < bound$} \EndIf}\label{jd-line}
		\State{\textbf{if} $newCost < edge$.dst.cost \textbf{then}}\label{relax:start}
		\State{$\quad \ edge$.dst.cost $\gets newCost$}
		\State{$\quad \ edge$.dst.predecessor $\gets node$}
		\State{$\quad \ Q$.add($edge$.dst)}\label{relax:end}
		\EndFor
		\EndWhile
		
		\EndFunction
		\State{\Return \textsc{null}}
	\end{algorithmic}
	\caption{Pseudo-code of the Dijkstra algorithm and the BD adaptation (shown in light red).
		Note that, depending on the data structure in use (see Sec.~\ref{sec:ds}) for the priority queue ($Q$), the pseudo-code may vary slightly.
		We show here the most common pseudo-code using a heap, which we used for our implementation and which performs best among the available data structures~\cite{cherkassky1996shortest}.}
	\label{pseudo-code}
\end{figure}

For an SP search, the Dijkstra algorithm~\cite{dijkstra1959note} discovers paths in increasing order of cost from the source node and stops once the destination is reached.
The pseudo-code is shown in Fig.~\ref{pseudo-code}.
When a bound is provided to Dijkstra (or BD), two different cases can happen.

\begin{figure}
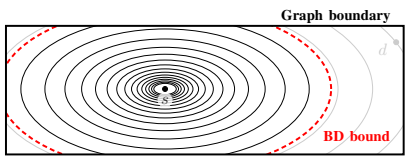

	\centering
	\includestandalone[width=0.75\linewidth]{images/sp-tree-search-jd}
	\caption{
		%As elaborated in Sec.~\ref{sec:impact-distance}, as the nodes get closer to each other compared to the graph size (i.e., as the size of the graph boundary rectangle increases), BD can more often stop the expansion of Dijkstra before the graph boundary itself does it.
		%Illustration of the breadth-first search of the Dijkstra algorithm for an SP search.
		Dijkstra discovers paths in increasing order of cost (as illustrated by the ellipses representing nodes which are equidistant from the source node) and stops once the destination ($d$) is reached.
		BD (dashed ellipse) terminates the search at a given cost from the source node, once it is known that longer paths will not be considered by the overlay algorithm.
		%As elaborated in Sec.~\ref{sec:sp}, BD allows to avoid the exploration of unnecessary areas of the network and hence to reduce runtime.
	}\label{fig:sp-search-jd-distance}
\end{figure}

\subsubsection{BD: The Destination is Further than the Bound}\label{sec:sp-jd-infeasible}

First, the provided bound can be lower than the cost of the SP to the destination.
In this case, BD avoids exploring unnecessary parts of the network (Fig.~\ref{fig:sp-search-jd-distance}) by preventing Dijkstra from considering paths with a cost greater than the provided bound and hence terminating before reaching the destination, i.e., before Dijkstra would have terminated.
The path returned is then ``\textsc{NULL}'', which is considered by the overlay algorithm in the same way as a path which is too costly: it does not use it.
\subsubsection{BD: The Destination is Closer than the Bound}\label{sec:sp-jd-feasible}

Second, the provided bound can be greater than the cost of the SP to the destination.
In this case, BD might appear useless, as it will, like Dijkstra, terminate when reaching the destination.
However, BD can also have a benefit in this case.
Indeed, BD can avoid putting an element in the queue whose associated cost is greater than the provided bound (line~\ref{jd-line} in Fig.~\ref{pseudo-code}).
Because such elements can be reached before the destination, this allows BD to avoid unnecessary operations (lines~\ref{relax:start}-\ref{relax:end}) and to reduce the size of its priority queue, thereby increasing the speed of the upcoming popping operations (line~\ref{popping}).
Let us consider an example where the cost of the SP to the destination is $16$ and the BD bound is $18$.
BD pops an element with an associated cost value of $15$ out of its queue (line~\ref{popping}) and expands it.
This expansion leads to elements with associated costs $20$, $22$ and $24$.
While the traditional Dijkstra would execute lines~\ref{relax:start}-\ref{relax:end}, BD knows that these elements will never be used.
Hence, BD can directly discard these elements (line~\ref{jd-line}), thereby preventing executing lines~\ref{relax:start}-\ref{relax:end}.
As a result, even when the bound provided to BD is greater than the cost of the SP to the destination, BD can be beneficial.
We will confirm this in our evaluations (Sec.~\ref{sec:ldp-eval}).
However, the benefit is expected to decrease as the bound gets greater.
Indeed, the number of elements that can be discarded will decrease.
In the example above, a higher BD bound of $23$ would for example allow to discard only one of the elements, rather than the three of them with a BD bound $18$.
Note that this phenomenon also happens when the destination is further than the provided bound (Sec.~\ref{sec:sp-jd-infeasible}).

\subsection{Impact on an SPT Search}\label{sec:sp-tree}

For an SPT search, Dijkstra behaves as in the SP case (Fig.~\ref{pseudo-code}) but instead of stopping when reaching a given node (line~\ref{single-destination} of Fig.~\ref{pseudo-code}), the algorithm stops when its priority queue is empty, i.e., when it has reached all the nodes.

When provided with a bound, BD can potentially stop the SPT expansion before exploring the whole graph.
That is, BD can avoid waiting for reaching some nodes which are too far away.
As for the single-destination case, this allows to reduce the runtime by preventing the exploration of unnecessary parts of the network (see Sec.~\ref{sec:sp-jd-infeasible}) and by avoiding unnecessary operations and the addition of unnecessary elements to the priority queue (see Sec.~\ref{sec:sp-jd-feasible}).

\subsection{BD Idea for Bellman-Ford (BF)}\label{sec:bf}

The \emph{Bellman-Ford} (BF) algorithm~\cite{bellman1958routing, ford1956network} is another algorithm for solving SP/SPT problems.
Because of its structure, the algorithm is more often used in distributed implementations.
However, it is also used as a subroutine of other complex algorithms where Dijkstra cannot be used (e.g., DEB, see Sec.~\ref{sec:jd-bf}).
While the structure of the BF algorithm is very different from the structure of Dijkstra, it also discovers paths in increasing order of cost.
Hence, the BD idea can also be applied to BF by simply discarding paths more costly than the given bound.
As a result, algorithms making use of the BF algorithm as a subroutine can also apply the BD principle.

\subsection{BD Idea for Chong's Algorithm}\label{sec:chong}

The problem of finding the \emph{k shortest paths} (kSP) between two nodes (or the kSPT from one node to several destinations) also arises often as a subroutine of more complex algorithms (e.g., kDCBF, see Sec.~\ref{sec:jd-sp-tree-also}).
\emph{Chong's algorithm}~\cite{chong1995finding} solves this problem by assuming that the $k$ value is known a priori.
The algorithm is identical to the Dijkstra algorithm but keeps track, at each node, instead of one single path, of the current $k$ best paths found.
Hence, the BD idea can be applied to Chong's algorithm in the same way as it is applied to Dijkstra.
As a result, algorithms making use of Chong's algorithm as a subroutine can also apply the BD principle.
%can also prevent the exploration of unnecessary parts of the network by discarding paths longer than the given bound.
BF can also be adapted to a static $k$SP algorithm by also simply keeping track of the current $k$ best paths found towards each node.
This adaptation can also apply the BD principle.

\section{Application: BD for CSP Routing}\label{sec:csp}

\renewcommand{\arraystretch}{1.35}
\begin{table}[t] \centering
	\footnotesize
	\begin{tabular}[width=\linewidth]{|c|cccc|c|c|}
		\multicolumn{1}{c}{} & \multicolumn{6}{c}{\textsc{Number of BD usages}} \\
		\hline
		\multirow{2}{*}{Algorithm} & \multicolumn{4}{ |c| }{Delay} & Cost & Comb. \\
		& BF & SP & SPT & kSPT & SPT & SPT \\
		\hline
		\hline
		\multicolumn{7}{|l|}{\emph{Algorithms that Cannot Use BD} (Sec.~\ref{sec:no-jd})} \\
		\hline
		\hline
		\underline{CBF}~\cite{widyono1994design} & & & & & & \\
		\underline{A*Prune}~\cite{liu2001prune} & & & & & & \\
		\underline{kSPMC}~\cite{guck2017unicast} & & & & & & \\
		\underline{E\_MCOP}~\cite{gang2002heuristic} & & & & & & \\
		SMS-PBO~\cite{sriram1998preferred} & & & & &&  \\
		kLARAC~\cite{jia2001heuristic} & & & & & & \\
		\hline
		\hline
		\multicolumn{7}{|l|}{\emph{Algorithms that Can Use BD for SP Only} (Sec.~\ref{sec:jd-sp-only})} \\
		\hline
		\hline
		LDP~\cite{guck2017unicast} & & 1 & & & & \\
		FB~\cite{lee1995routing} & & (0, 1) & & & & \\
		LARAC~\cite{aneja1978constrained, handler1980dual, blokh1996approximate, juttner2001lagrange} & & 1 & & & & \\
		\underline{LARACGC}~\cite{handler1980dual} & & 1 & & & & \\
		\underline{SCRC}~\cite{santos2007improved} & & 1 & & & & \\
		DCCR~\cite{guo2003search} & & 1 & & & & \\
		SSR+DCCR~\cite{guo2003search} & & 1 & & & & \\
		\hline
		\hline
		\multicolumn{7}{|l|}{\emph{Algorithms that Can Use BD for SPT} (Sec.~\ref{sec:jd-sp-tree-also})} \\
		\hline
		\hline
		DCUR~\cite{salama1997distributed, reeves2000distributed} & & & 1 & & (0, 1) & \\
		SF-DCLC~\cite{liu2005efficient} & & & 1 & & (0, 1) & \\
		SMS-CDP~\cite{sriram1998preferred} & & & 1 & & (0, 1) & \\
		SMS-RDM~\cite{sriram1998preferred} & & & 1 & & & \\
		IAK~\cite{ishida1998delay} & & & 1 & & & \\
		DCR~\cite{sun1998new} & & 1 & & & (0, 1) & \\
		H\_MCOP~\cite{korkmaz2001multi} & & & 1 & & &  \\
		kH\_MCOP~\cite{korkmaz2001multi} & & & 1 & & & \\
		NR\_DCLC~\cite{feng2002performance} & & (0, 1) & & & & $\geq 0$ \\
		MH\_MCOP~\cite{gang2002heuristic} & & & 1 & & & $\geq 0$ \\
		DCBF~\cite{jia2001heuristic} & & & 1 & & & \\
		kDCBF~\cite{jia2001heuristic} & & & & 1 & & \\
		\hline
		\hline
		\multicolumn{7}{|l|}{\emph{Algorithm that Can Use BD for BF} (Sec.~\ref{sec:jd-bf})} \\
		\hline
		\hline
		DEB~\cite{cheng2003new} & 1 & & & & &  \\
		\hline
	\end{tabular}
	\caption{Number of times some constrained shortest path (CSP) algorithms can make use of BD based on the metric (cost, delay or a combination) and algorithm on which BD can be applied (SP/SPT refers to Dijkstra, kSPT to Chong and BF to Bellman-Ford for SP).
		When the number of times BD can be used depends on the routing request, the set of possible values is given between parentheses and unbounded values are given using the $\geq$ symbol.
		\underline{Underlined} algorithms are optimal.}\label{tab:algorithms}
\end{table}

In this section, we show that BD can be used by a wide range of algorithms by showing \emph{(i)} how existing \emph{constrained shortest path} (CSP) algorithms can replace their SP and SPT subroutines with BD, and \emph{(ii)} how bounds can be provided to these BD runs.

The CSP problem consists in finding the shortest path (in terms of a first metric referred to as \emph{cost}) such that a second metric (referred to as \emph{delay}) is lower than a given bound.
CSP algorithms use SP/SPT subroutines using either the cost metric, the delay metric, or a combination of both~\cite{guck2017unicast} as optimization metric.
Tab.~\ref{tab:algorithms} shows, for each algorithm, how many times it can replace a SP/SPT run with a BD run.
The cases for which BD can be used are separated based on the metric (cost, delay or a combination) and on the algorithm on which the BD principle is applied (SP/SPT refers to Dijkstra, kSPT to Chong and BF to Bellman-Ford for SP).

In a recent survey~\cite{guck2017unicast}, Guck \emph{et al.} presented 26 different CSP algorithms, out of which only 6 cannot make use of BD.
In the following sections (and in Tab.~\ref{tab:algorithms}), algorithms are referred to using their acronyms as defined in~\cite{guck2017unicast}.
%Due to space constraints, the original references of the different algorithms are not given.
%These can be found in~\cite{guck2017unicast}.

\subsection{Algorithms that Cannot Use BD}\label{sec:no-jd}

First, \emph{CBF}~\cite{widyono1994design}, \emph{A*Prune}~\cite{liu2001prune} and \emph{SMS-PBO}~\cite{sriram1998preferred} have a specific structure making use of no underlying (k)SP/SPT algorithm and can hence not make use of BD.
Second, \emph{kSPMC}~\cite{guck2017unicast}, \emph{E\_MCOP}~\cite{gang2002heuristic} and \emph{kLARAC}\cite{jia2001heuristic} exclusively make use of kSP and SP algorithms to which no bound can be provided.
%They can hence also not make use of BD.

\subsection{Algorithms that Can Use BD for SP Only}\label{sec:jd-sp-only}

The \emph{LDP}~\cite{guck2017unicast}, \emph{FB}~\cite{lee1995routing}, \emph{LARAC}~\cite{aneja1978constrained, handler1980dual, blokh1996approximate, juttner2001lagrange}, \emph{LARACGC}~\cite{handler1980dual}, \emph{SCRC}~\cite{santos2007improved}, \emph{DCCR}~\cite{guo2003search}, and \emph{SSR+DCCR}~\cite{guo2003search} algorithms run a least-delay SP procedure (i.e., optimizing the delay metric) which can make use of BD by using the bound of the original problem.
After this least-delay SP run, the \emph{LARAC}, \emph{LARACGC}, \emph{SCRC}, \emph{DCCR}, and \emph{SSR+DCCR} algorithms run one or several least-cost SP runs (i.e., optimizing the cost metric).
These runs could be provided with the cost of the least-delay path as bound. 
However, if provided with this bound, this BD run will always be in the case described in Sec.~\ref{sec:sp-jd-infeasible} where the provided bound is greater than the cost of the shortest path to the destination.
As we will see in Sec.~\ref{sec:ldp-eval}, on average, the usage of BD in such a case increases the runtime of the SP run.
As a result, we do not consider the least-cost run as a BD run.
%The \emph{LARACGC}, \emph{SCRC}, \emph{DCCR}, and \emph{SSR+DCCR} further execute a $k$SP run to which no bound can be provided.

\subsection{Algorithms that Can Use BD for SPT}\label{sec:jd-sp-tree-also}

The \emph{DCUR}~\cite{salama1997distributed, reeves2000distributed}, \emph{SF-DCLC}~\cite{liu2005efficient}, \emph{SMS-CDP}~\cite{sriram1998preferred}, \emph{SMS-RDM}~\cite{sriram1998preferred}, \emph{IAK}~\cite{ishida1998delay}, \emph{DCR}~\cite{sun1998new}, \emph{H\_MCOP}~\cite{korkmaz2001multi}, \emph{kH\_MCOP}~\cite{korkmaz2001multi}, \emph{DCBF}~\cite{jia2001heuristic} and \emph{kDCBF}~\cite{jia2001heuristic} algorithms run a least-delay search to which the delay bound of the CSP problem can be provided as a bound.
While \emph{DCR} runs a least-delay SP search and \emph{kDCBF} a least-delay kSPT (Chong) search, all the others run a least-delay SPT search.
DCUR, SF-DCLC, SMS-CDP and DCR then possibly execute a least-cost SPT run to which the cost of the least-delay path from the source to the destination can be provided as a bound.
Indeed, any path with a cost higher than the least-delay path will never be used by the algorithms, as they would then rather choose to follow the least-delay path, which has both a lower cost and delay.
The \emph{IAK}, \emph{H\_MCOP}, \emph{kH\_MCOP}, \emph{DCBF} and \emph{kDCBF} algorithms further execute a least-cost (k)SP search.
As for the algorithms in Sec.~\ref{sec:jd-sp-only}, a bound could be provided to this least-cost run but, for the same reason, we do not consider it.

The \emph{NR\_DCLC} algorithm~\cite{feng2002performance} starts like FB and can hence make use of BD in the same way.
Then, if the problem is feasible, it runs several times H\_MCP~\cite{korkmaz2001multi, gang2002heuristic} (an MCP algorithm), a modified version of H\_MCOP, to improve on the least-delay path result.
H\_MCP uses a metric combining the cost and delay metrics for its SPT search.
Since bounds on both the delay (the bound of the CSP problem) and on the cost (the cost of the best path found so far) are known, the first step of H\_MCP can also make use of BD.
Hence, NR\_DCLC can further make use of BD by using H\_MCP with BD.

The \emph{MH\_MCOP} algorithm~\cite{gang2002heuristic} is similar to NR\_DCLC but, instead of using H\_MCP to improve on the least-delay path result, H\_MCP is used to improve on the path found by H\_MCOP.
Hence, MH\_MCOP can make use of BD by using both H\_MCOP and H\_MCP with BD.	

\subsection{Algorithm that Can Use BD for BF}\label{sec:jd-bf}

The \emph{DEB} algorithm~\cite{cheng2003new} runs a least-cost and a least-delay SP search using BF.
As for \emph{LARAC}, a bound can be provided to both the least-delay and least-cost searches but we only consider the least-delay search as a BD run.

\section{Evaluation}\label{sec:evaluation}

The goal of our evaluation is twofold.
First, in order to confirm our expectations of Sec.~\ref{sec:sp} and \ref{sec:sp-tree}, we quantify the impact of BD on an SP and an SPT run.
To do so, we observe the behavior of the LDP (Sec.~\ref{sec:ldp-eval}) and IAK (Sec.~\ref{sec:iak-eval}) algorithms, which are using BD respectively for a single SP and a single SPT run based on the delay metric. 
Second, in order to confirm the applicability of BD, in Sec.~\ref{eval:all-csp}, we observe its impact on the performance of the CSP algorithms presented in Sec.~\ref{sec:csp}.
Because of the big amount of resulting data, we only present here the most insightful and representative results.
The complete data and set of graphs has been made publicly available at \textit{https://lora.lkn.ei.tum.de}~\cite{heaven-page}.

Among all the runs performed during the evaluation, the paths returned by the algorithms with and without BD were always identical, thereby confirming that BD does not impact the output of the algorithms.
Hence, in the following, we only discuss the runtime of the algorithms.

The algorithms have been implemented using Java 8 and evaluated on an Ubuntu 16.04 PC equipped with an Intel Core i7-4790 CPU @ 3.60GHz.

\subsection{Setup}

In this section, we define the three dimensions (Sec.~\ref{dim-1} to \ref{dim-3}) along which we run our evaluation and describe how our plots (Sec.~\ref{sec:plots}) and routing requests (Sec.~\ref{sec:request}) are generated.

\subsubsection{First Dimension: Distance between Nodes}\label{dim-1}

From Fig.~\ref{fig:sp-search-jd-distance}, we can expect that, if the source and destination nodes are far apart from each other, the impact of BD will be lower.
Indeed, in most directions, the graph boundary itself will be expected to stop the expansion of Dijkstra before BD does it.
If the source and destination nodes are closer to each other compared to the graph size (alternatively, if the graph boundary rectangle in Fig.~\ref{fig:sp-search-jd-distance} gets bigger), we can expect that the BD bound will be reached more often before the boundary of the network and hence BD will provide more benefit.
That is, the impact of BD potentially depends on the relative distance (in terms of cost) between the source and destination nodes compared to the size of the topology.
To ease the definition of this dimension, we only consider a grid topology of size $N\times N$.
For a given grid size, we define 10 different so-called \emph{distance buckets}.
These buckets correspond to source and destination nodes pairs whose least-hop distance in the grid is between 0 and 10\%, 10\% and 20\%, ..., 90\% and 100\% of the longest path in the grid (i.e., of $2 \times (N -1)$).

\subsubsection{Second Dimension: Tightness of the Constraint}\label{dim-2}

%We also observed in Sec.~\ref{sec:jd} that the impact of BD potentially differs based on the tightness of the constraint of the overlay problem (CSP problem in our case).
The delay constraint can range from a loose value for which the least-cost path is feasible to tight values for which the problem is infeasible.
Within this range, we define 7 ranges of equal sizes.
We refer to these ranges as \emph{delay levels}.

\subsubsection{Third Dimension: Grid Size}\label{dim-3}

We further consider the grid size $N$ as an evaluation dimension.
We vary $N$ from $6$ to $20$.
We observed that the grid size does not influence the impact of BD.
Hence, we here ignore this dimension (we always aggregate all the results for all the sizes) but the corresponding graphs are available online~\cite{heaven-page}.

\subsubsection{Plots Generation}\label{sec:plots}

For each algorithm, we generate plots showing the distribution\footnote{After removing the values below the 1\% percentile and above the 99\% percentile, the distributions are shown as boxplots showing the 10\%, 25\%, 50\%, 75\% and 90\% percentiles. A red square identifies the average. Versions of the plots also showing the outliers are available on the accompanying web interface~\cite{heaven-page}.} of the \emph{runtime ratios} observed for the different values of a given dimension (i.e., node distance, delay level or grid size).
For a given request, the runtime ratio is defined as the runtime of the algorithm without BD divided by the runtime of the algorithm with BD.
%Hence, values greater than 1 corresponds to a situation in which BD reduces the runtime of the algorithm.
The hidden dimensions are either aggregated (i.e., the runtime ratios for all their values are incorporated in the distributions) or only a specific value of these dimensions is incorporated in the distributions.% (e.g., only values of one of the distance buckets).

\subsubsection{Requests Generation}\label{sec:request}

For each algorithm and for each combination of distance bucket, delay level and grid size, we generate random cost and delay values between $1$ and $2$ for each link and we randomly generate 5000 requests (within the corresponding distance bucket and delay level).
The 5000 requests are then solved by the considered algorithm and its corresponding version with BD.
The first 500 runs are used as warm-up for the Java HotSpot optimizer and their results are not considered.
The order in which the algorithm and its BD version are run is alternating.
This prevents the Java HotSpot optimizer from optimizing one of the run over the other.
The distance bucket dimension cannot be aggregated simply by considering all the runtime ratios for all the different buckets.
Indeed, considering random source-destination pairs, small distances are more probable than long distances.
Hence, for plots aggregating the distance bucket dimension, for each algorithm and combination of delay level and grid size, we generate 50000 (out of which 5000 are used as warm-up) requests by randomly selecting a source and a destination node.
% \footnote{10 times more than for each distance bucket since there are 10 different distance buckets.}

\subsection{LDP: Influence of BD on an SP Search}\label{sec:ldp-eval}

We first observe the impact of BD on the runtime of LDP.
This allows us to gain insight into the behavior of BD for SP runs (see Sec.~\ref{sec:sp}).

\begin{figure}
	\centering
	\includegraphics[width=0.75\linewidth]{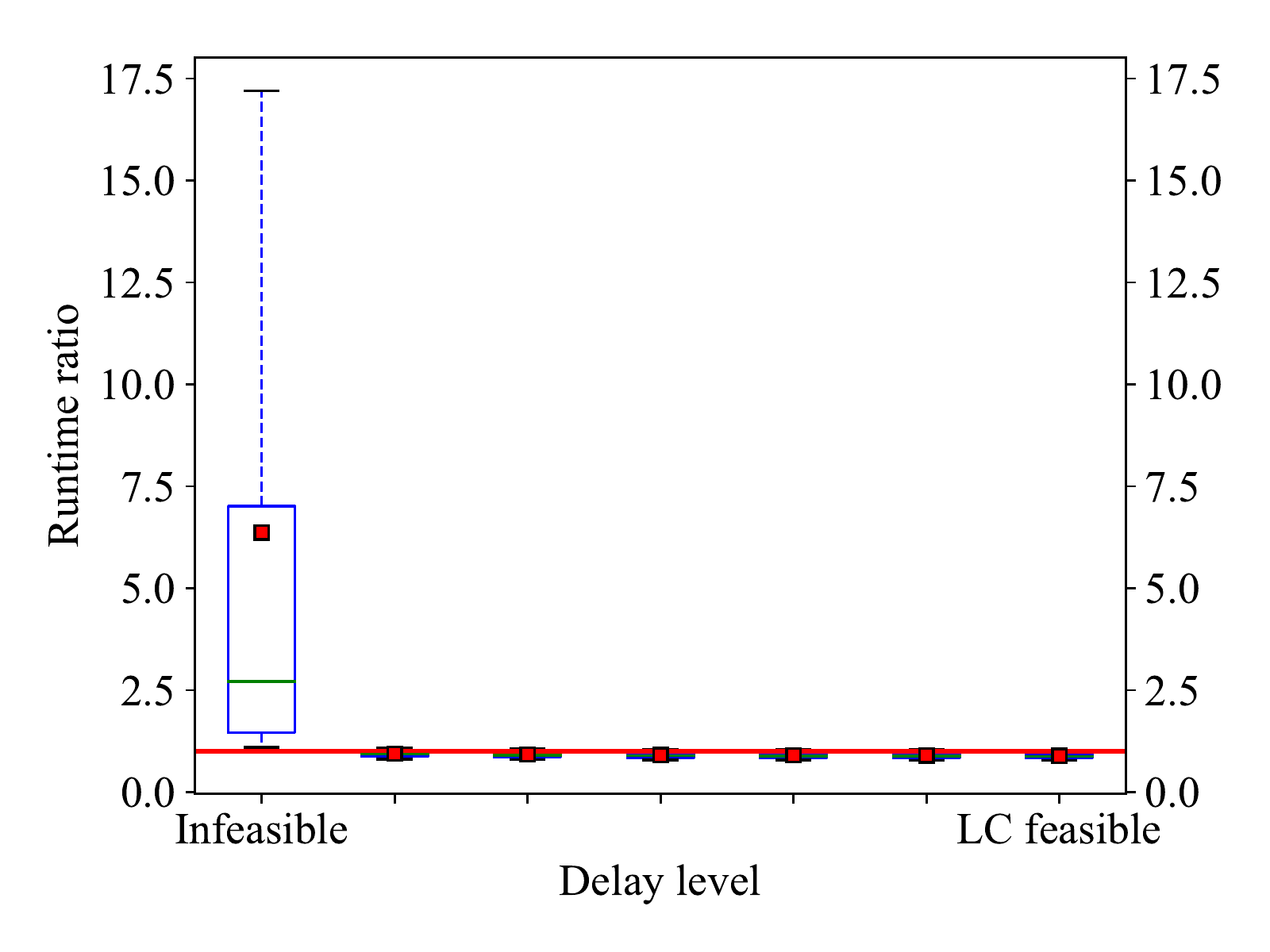}
	\caption{Runtime ratios of LDP for different delay levels.
		On average, for single-destination SP searches, BD is only useful when the provided bound is lower than the cost of the shortest path to the destination.}\label{fig:ldp-average-feasible-useless}
\end{figure}

Fig.~\ref{fig:ldp-average-feasible-useless} shows the impact of BD on the runtime of LDP for the different delay levels, all the other dimensions being aggregated.
As expected, we observe that BD allows to dramatically reduce (more than 6 times faster on average) the runtime of an SP search when the bound is lower than the cost of the shortest path to the destination (\emph{infeasible} delay level, corresponding to the scenario described in Sec.~\ref{sec:sp-jd-infeasible}).
For all the other cases (the delay bound is greater than the cost of the shortest path to the destination -- see Sec.~\ref{sec:sp-jd-feasible}), we however observe that, \emph{on average}, the additional runtime induced by BD for checking if the bound is violated (line~\ref{jd-line} in Fig.~\ref{pseudo-code}) is not compensated by its benefit.
Indeed, we observe that the runtime ratios are, on average, slightly lower than 1.

\begin{figure}
	\centering
	\includegraphics[width=0.75\linewidth]{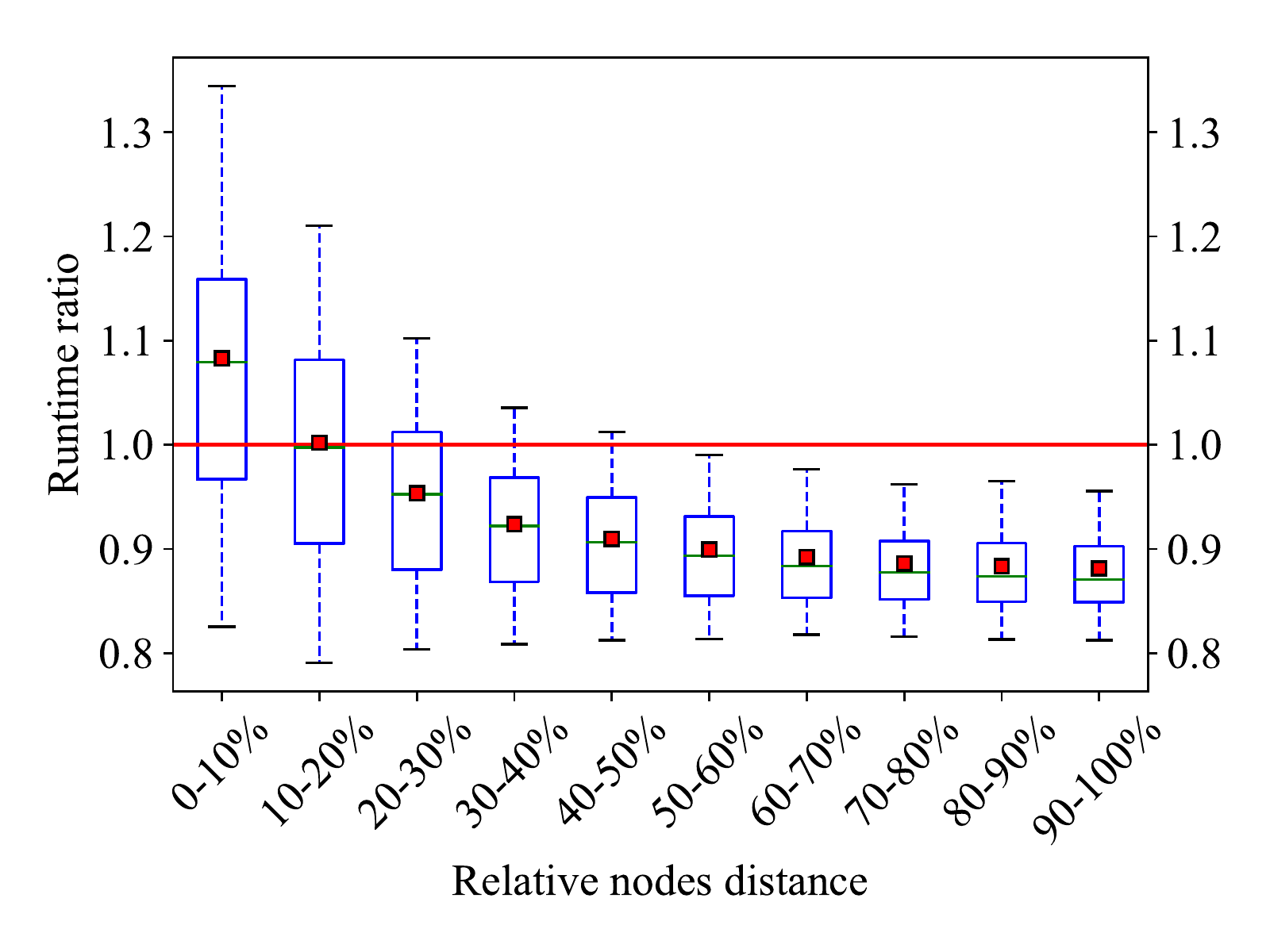}
	\caption{Runtime ratios of LDP for the different distance buckets and for the first delay level that is feasible.
		For SP searches, in some favorable cases (short distances), BD can still be beneficial even if the provided bound is greater than the cost of the shortest path to the destination.}\label{fig:ldp-feasible-still-useful-but-gets-worst-if-big}
\end{figure}

However, interestingly, even if the provided bound is greater than the cost of the shortest path to the destination, there are cases for which BD reduces runtime (this is due to the fact that BD then avoids to place unnecessary elements in the priority queue -- see. Sec.~\ref{sec:sp-jd-feasible}).
Fig.~\ref{fig:ldp-feasible-still-useful-but-gets-worst-if-big} shows the impact of BD on the runtime of LDP for the different distance buckets, the grid size dimension being aggregated and for the first feasible delay level.
The figure confirms that BD can also improve the runtime of an SP search when the provided bound is greater than the shortest path to the destination (see Sec.~\ref{sec:sp-jd-feasible}) but that the benefit of BD only balances its additional overhead for low distances and tight bounds.
Fig.~\ref{fig:ldp-average-feasible-useless} however shows that, \emph{on average}, BD is only beneficial when the provided bound is lower than the cost of the shortest path to the destination. 

As expected in Sec.~\ref{dim-1}, Fig.~\ref{fig:ldp-feasible-still-useful-but-gets-worst-if-big} also shows that, when the provided bound is greater than the shortest path to the destination, the impact of BD decreases as the distance between the nodes compared to the topology size increases.
When the provided bound is lower than the shortest path to the destination, this effect is compensated by the fact that, when the distance is low, Dijkstra will anyway terminate before BD can stop it, thereby preventing BD from significantly reducing the search space.
Hence, in this case, the impact of BD is relatively stable along the different distance buckets.
This can be seen on the additional graphs available online~\cite{heaven-page}.

\subsection{IAK: Influence of BD on an SPT Search}\label{sec:iak-eval}

We observe the impact of BD on the runtime of IAK.
As IAK simply runs a least-cost SP search (without BD) and a least-delay SPT search with BD, this allows us to gain insight into the behavior of BD for SPT runs.

\begin{figure}
	\centering
	\includegraphics[width=0.75\linewidth]{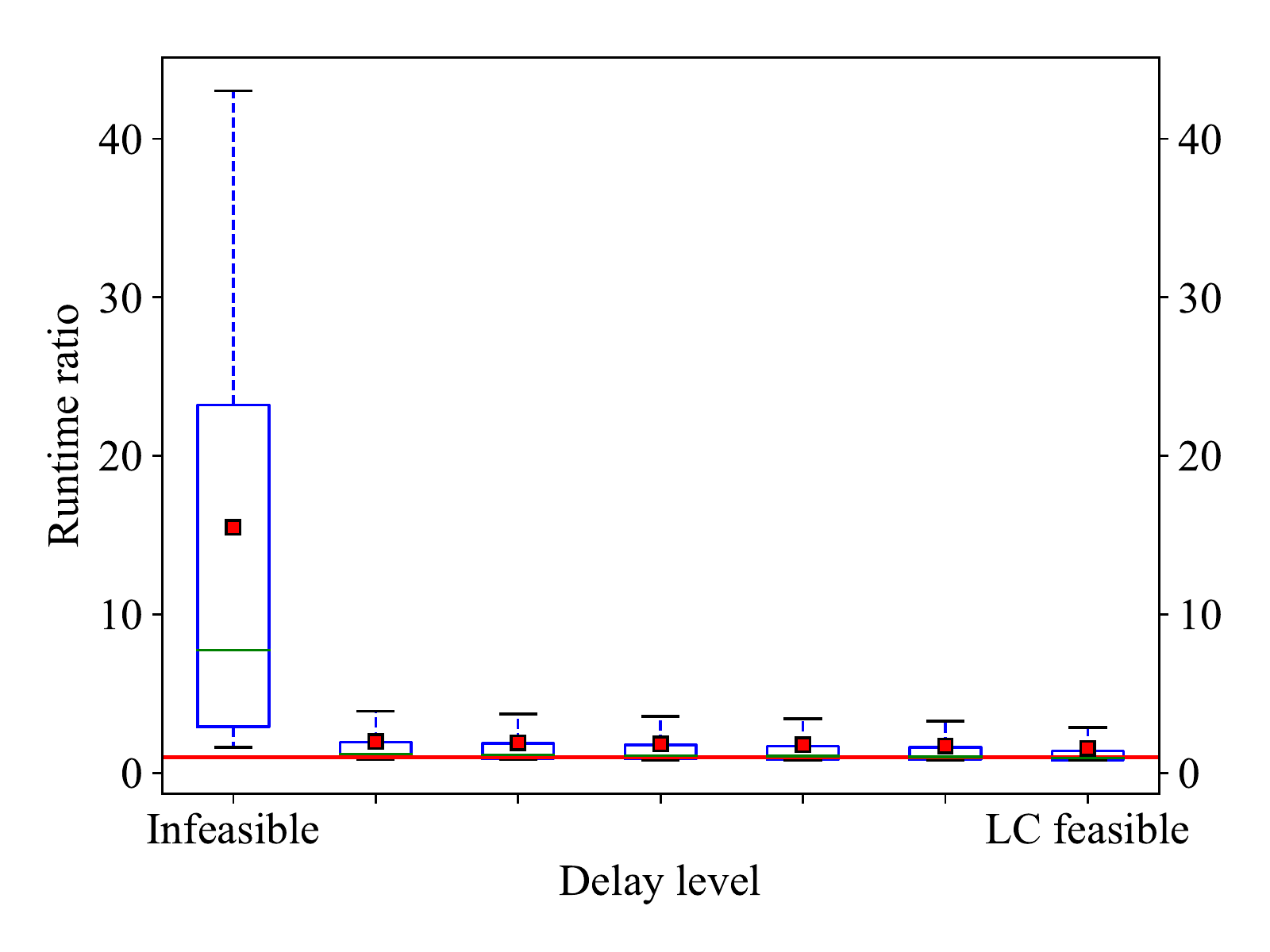}
	\caption{Runtime ratios of IAK for different delay levels.
		For SPT searches, BD is beneficial in any case but better when the provided bound is lower.}\label{fig:iak-delay-level}
\end{figure}

Fig.~\ref{fig:iak-delay-level} shows the impact of BD on the runtime of IAK for the different delay levels, the other dimensions being aggregated.
In comparison to Fig.~\ref{fig:ldp-average-feasible-useless}, this shows that BD has a much higher impact for SPT runs (up to 15 times faster on average for infeasible cases, against 6 times faster for SP searches).
Further, we observe that, even when the problem is feasible, BD still provides benefit.
This is because, even if the destination is closer than the provided bound, other nodes further away may be neglected by BD.
Hence, this shows that, on average, BD is useful in any case for SPT searches.
As expected, we observe that the impact of BD decreases as the delay bound gets looser.

In further evaluations available online~\cite{heaven-page}, we have observed that, as for SP runs, the impact of BD on SPT searches decreases as the distance between the source and destination nodes of the original CSP problem increases.

%\begin{figure}
%	\centering
%	\includegraphics[width=0.7\linewidth]{images/no-outliers-IAK-A-A-X}
%	\caption{Runtime ratios of IAK for the different distance buckets.
%		This shows that, for SPT searches, BD is beneficial in any case but better when the source and destination of the CSP problem are close to each other.
%		Though not shown in this graph, this behavior is common to most algorithms.}\label{fig:iak-distance}
%\end{figure}

% 	Fig.~\ref{fig:iak-distance} shows the impact of BD on the runtime of IAK for the different distance buckets, the other dimensions being aggregated.
% 	We observe that BD has less impact when the source and destination nodes of the CSP problem are further away from each other.
% 	This was expected.
% 	Indeed, as nodes get further away from each other, the SPT search is more often blocked by the graph boundary itself rather than by the BD bound.
% 	In additional evaluations for the other algorithms (available online~\cite{heaven-page}), we observed that this behavior is common to most algorithms.
% 	As Fig.~\ref{fig:iak-delay-level}, Fig.~\ref{fig:iak-distance} also confirms that BD is more efficient for SPT runs than for single-destination SP runs and that it is, on average, beneficial.      	  

\subsection{BD Impact on All CSP Algorithms}\label{eval:all-csp}

In this section, we observe the impact of BD on all the CSP algorithms presented in Sec.~\ref{sec:csp}.
Algorithms requiring parameters have been configured as in \cite{guck2017unicast}.
In the plots, the parameters values are appended to the algorithm names.

Because too slow, SCRC, LARACGC, SMS-CDP, SMS-RDM, SMS-PBO and DEB were not able to run the evaluation in a reasonable amount of time.
However, because of their similarity with LARAC, the impact of BD on LARACGC and SCRC is supposed to be similar to the impact on LARAC.

\begin{figure}
	\centering
	\includegraphics[width=0.95\linewidth]{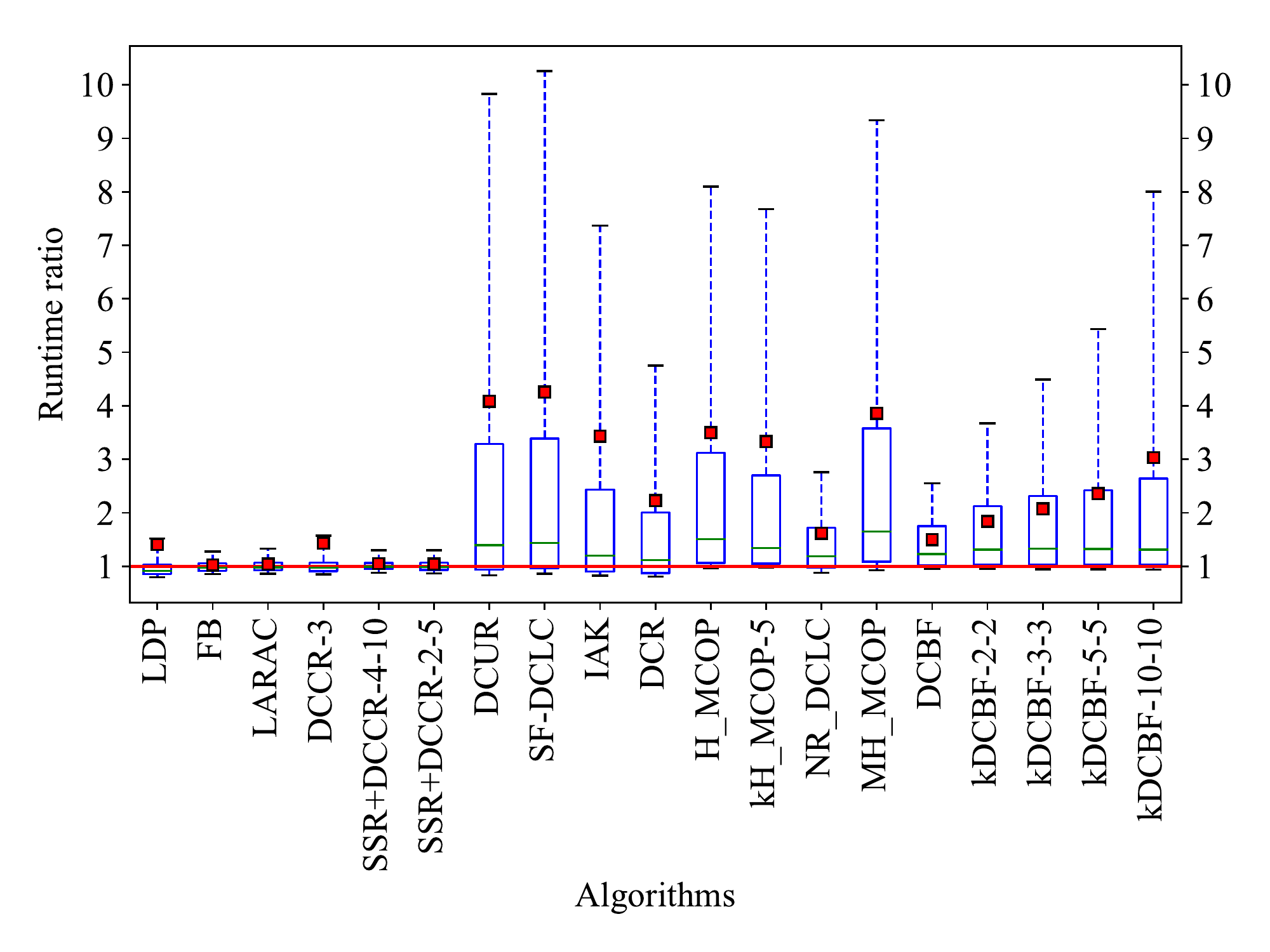}
	\caption{Distributions of the runtime ratios of the different algorithms, all the dimensions being aggregated.
		We observe that BD can greatly reduce the runtime of some algorithms (up to 4 times faster, i.e., runtime reduced by up to 75\% on average for some algorithms).}\label{fig:ALL}
\end{figure}

Fig.~\ref{fig:ALL} shows the runtime ratios of all the algorithms, all the dimensions being aggregated.
As can be seen, BD is, on average, beneficial for all the algorithms.
However, we can see that algorithms which can only use BD for SP runs (LDP, FB, LARAC, DCCR and SSR+DCCR -- Sec.~\ref{sec:jd-sp-only}) are only slightly improved.
This was expected based on our observations of Sec.~\ref{sec:ldp-eval}.
Indeed, the average impact of BD on SP runs is marginal.
On the other hand, we can see that the runtime of algorithms which can use BD for SPT runs (DCUR, SF-DCLC, IAK, DCR, H\_MCOP, kH\_MCOP, NR\_DCLC, MH\_MCOP, DCBF and kDCBF -- Sec.~\ref{sec:jd-sp-tree-also}) can be dramatically improved by BD.
For example, on average, DCUR, SF-DCLC and MH\_MCOP are 4 times faster with BD, i.e., their runtime is reduced by 75\% with BD.
Also, most algorithms see their runtime improved by at least 20\% in 50\% of the cases.
DCUR, SF-DCLC, IAK and DCR present an interesting behavior.
The two algorithms benefiting the most from BD are DCUR and SF-DCLC, because both their least-cost and least-delay SPT runs can use BD.
Then, IAK benefits less because its least-cost SP run cannot benefit from BD.
Finally, DCR benefits less than IAK even though its least-delay SP run and its least-cost SPT run can both benefit from BD.
This shows that, while SPT runs benefit more from BD than SP runs, least-delay SPT runs benefit more from BD than least-cost SPT runs.
MH\_MCOP further shows that using BD for SPT runs based on a combination of the cost and delay metrics can provide as much benefit as for SPT runs based on the delay metric solely.

\begin{figure}
	\centering
	\includegraphics[width=0.95\linewidth]{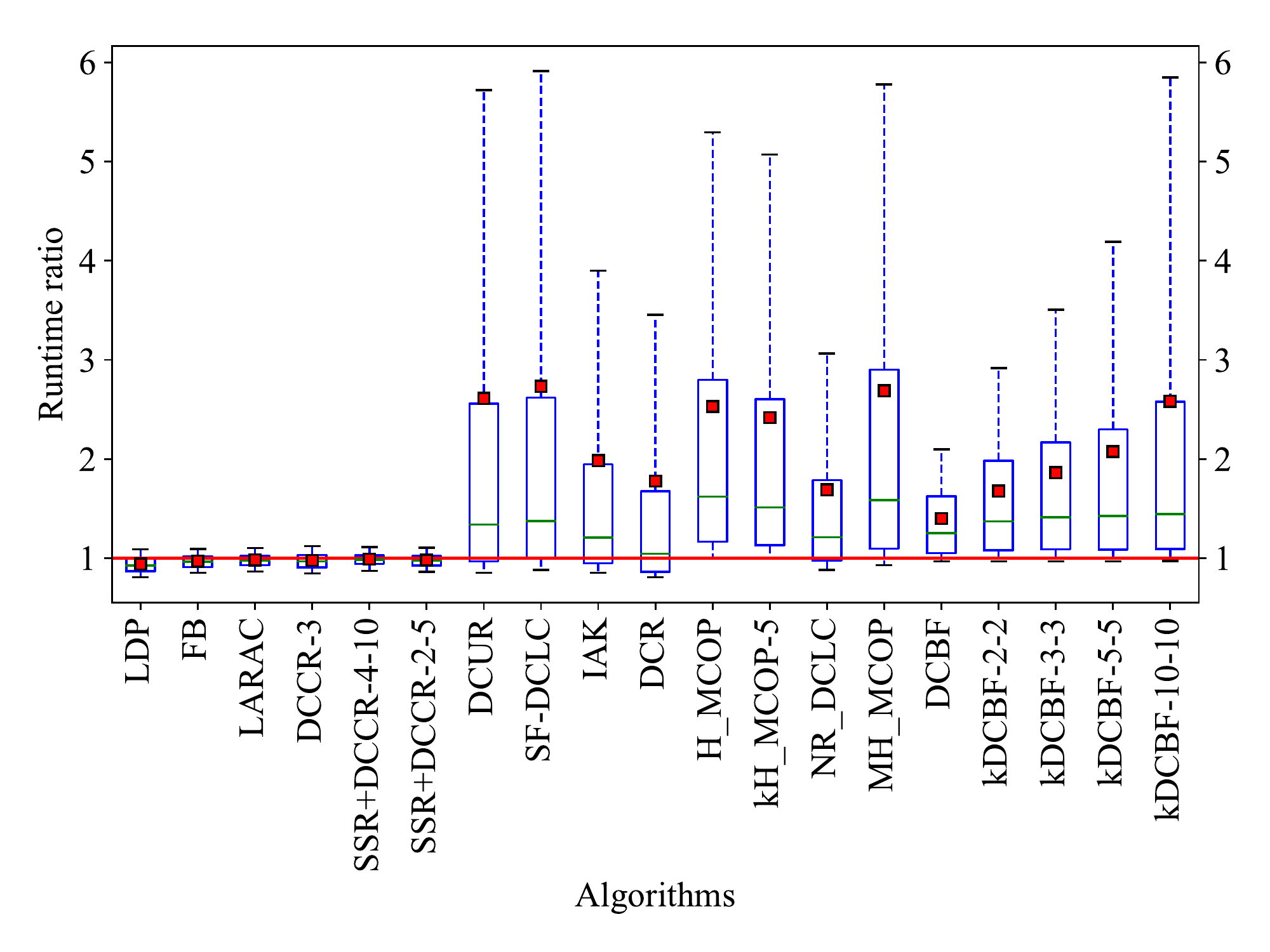}
	\caption{Distributions of the runtime ratios of the different algorithms for the first feasible delay level, the other dimensions being aggregated.
		We observe that, even outside of the infeasible case, BD can provide significant benefit to some algorithms (around $3$ times faster, i.e., runtime reduced by up to 66\% on average for some algorithms).}\label{fig:all-feasible-tight}
\end{figure}

In order to highlight that BD can also provide significant benefit for feasible delay bounds, Fig.~\ref{fig:all-feasible-tight} shows the distributions of the runtime ratios of the different algorithms for the first feasible delay level, the other dimensions being aggregated.
We can see that BD can still drastically reduce the runtime of some algorithms when the delay bound is feasible.
For example, DCUR, SF-DCLC, H\_MCOP, MH\_MCOP and kDCBF-10-10 are, in this case, on average around $3$ times faster with BD, i.e., their runtime is reduced by 66\%.
In further evaluations available online~\cite{heaven-page}, by aggregating all dimensions except the distance between the source and destination nodes, we have observed that, when the problem is not infeasible, BD is beneficial to the algorithms as long as the relative distance stays below 40-50\% of the topology size. 
When the nodes are further apart from each other, the graph boundary stops the expansion of Dijkstra before BD can have any significant impact.

\begin{figure}
	\centering
	\includegraphics[width=0.95\linewidth]{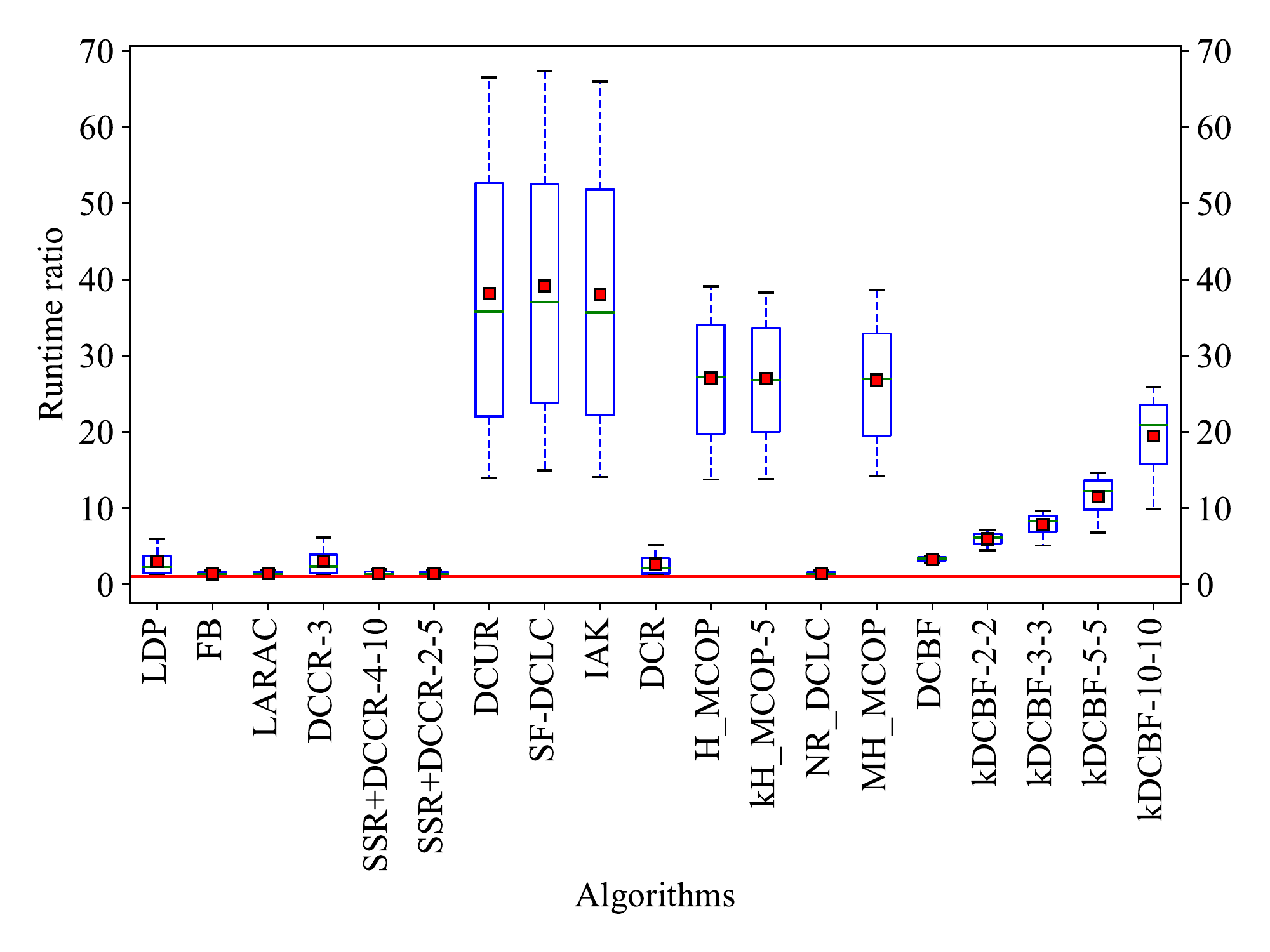}
	\caption{Distributions of the runtime ratios of the different algorithms for favorable cases (infeasible delay constraint and the 0-10\% distance bucket).
		We observe that, in these favorable cases, BD can drastically reduce the runtime of some algorithms (more than $25$ times faster, i.e., runtime reduced by at least 96\% on average for some algorithms).}\label{fig:all-infeasible-low-distance}
\end{figure}

We have seen that BD has potentially more impact when the delay constraint is tighter and the distance between the source and destination nodes is smaller.
Fig.~\ref{fig:all-infeasible-low-distance} shows the runtime ratios of all the algorithms for the infeasible delay level and for the 0-10\% distance bucket, the grid size dimension being aggregated.
We observe that, in this favorable case, BD allows to drastically reduce the runtime of all the algorithms, including those only using BD for SP runs.
For example, SF-DCLC, IAK, DCUR, H\_MCOP, kH\_MCOP and MH\_MCOP are more than $25$ times faster with BD, i.e., they see their runtime reduced by more than 96\% on average.
Interestingly, DCR and NR\_DCLC, which have a good average runtime improvement (see Fig.~\ref{fig:ALL}), do not benefit much in this favorable case.
This is because, in the infeasible case, both algorithms only run SP searches, thereby having a benefit similar to the algorithms only using BD for SP searches (e.g., LDP and LARAC).

\section{Conclusions}

\emph{Shortest path} (SP) and \emph{shortest paths tree} (SPT) algorithms are often used as subroutine of overlay algorithms solving more complex problems (e.g., the \emph{(multi-)constrained shortest path} (CSP and MCSP), the \emph{multi-constrained path} (MCP), and the \emph{constrained minimum Steiner tree} (CMST) problems).
In such a situation, it often happens that the result of an SP subroutine is not used if its total cost is greater than a given bound.
Because Dijkstra discovers path in increasing order of cost, we can terminate the execution of Dijkstra as soon as it reaches paths which have a cost greater than the known bound.
We refer to this adaptation of Dijkstra as \emph{bounded Dijkstra} (BD).
By terminating Dijkstra earlier, its search space is reduced, thereby reducing its runtime and hence the runtime of the overlay algorithm using it.
BD can be used by any routing algorithm making use of an underlying SP/SPT algorithm and that can provide a bound to this algorithm.
We evaluated the impact of BD on the specific example of CSP algorithms.
We have shown that BD does not impact the output of the algorithms but can dramatically decrease their runtime.
While BD can be beneficial for both SP and SPT searches, we showed that its benefit is greater for SPT runs. 
The runtime of some algorithms is reduced by 75\% on average.
We further showed that BD is more efficient for tight delay constraints and when the source and destination nodes of the CSP problem are close to each other compared to the size of the topology.
For these favorable cases, several algorithms see their runtime reduced by 96\% on average (i.e., BD allows them to be more than 25 times faster).

\section*{Acknowledgments}

This work has received funding from the European Union's Horizon 2020 research and innovation programme under grant agreement No.~671648 (VirtuWind) and from the German Research Foundation (DFG) under the grant numbers KE1863/6-1 and KE1863/8-1.

\balance

\bibliographystyle{IEEEtran}


\end{document}

%% file: images/sp-tree-search-jd.tex
\begin{tikzpicture}

 \coordinate (source) at (0, 0);
 \coordinate (destination) at (4.94, 1);

\draw[fill=black, draw=black] (destination) circle (1.5pt) node[left, xshift=-2pt, yshift=-4pt] {$d$};

\foreach \i in {1,...,18}
{
  \draw (source) ellipse (1.2^\i/15*3 and 1.2^\i/15*1.5);
}

\def\all{(-5.34,-2.7) rectangle (5.34,2.7)}
\def\jdellipse{(source) ellipse (1.2^15.4/14*3 and 1.2^15.4/14*1.5)}
\def\topology{(source) (-3.4,-1.4) rectangle (5.1,1.35);}

\begin{scope}[even odd rule]
  \clip \jdellipse \topology;
  \fill[white,opacity=0.8] \all;
\end{scope}

\draw[red, very thick, densely dashed] \jdellipse;

\begin{scope}[even odd rule]
  \clip \all \topology;
  \fill[white] \all;
\end{scope}

\node[red] at (4.1, -1) {\bfseries \small BD bound};

\draw[very thick, black] \topology;

\node at (3.65, 1.55) {\bfseries \small Graph boundary};

\draw[fill=black, draw=black] (source) circle (1.5pt) node[circle, below, fill=white, fill opacity=0.8, yshift=-2pt, inner sep= 1pt] {$s$};

\end{tikzpicture}